\PassOptionsToPackage{unicode}{hyperref}
\PassOptionsToPackage{hyphens}{url}
\PassOptionsToPackage{dvipsnames,svgnames,x11names}{xcolor}
\documentclass[
]{article}
\usepackage{amsmath,amssymb}
\usepackage{lmodern}
\usepackage{iftex}
\ifPDFTeX
  \usepackage[T1]{fontenc}
  \usepackage[utf8]{inputenc}
  \usepackage{textcomp} % provide euro and other symbols
\else % if luatex or xetex
  \usepackage{unicode-math}
  \defaultfontfeatures{Scale=MatchLowercase}
  \defaultfontfeatures[\rmfamily]{Ligatures=TeX,Scale=1}
\fi
\IfFileExists{upquote.sty}{\usepackage{upquote}}{}
\IfFileExists{microtype.sty}{% use microtype if available
  \usepackage[]{microtype}
  \UseMicrotypeSet[protrusion]{basicmath} % disable protrusion for tt fonts
}{}
\makeatletter
\@ifundefined{KOMAClassName}{% if non-KOMA class
  \IfFileExists{parskip.sty}{%
    \usepackage{parskip}
  }{% else
    \setlength{\parindent}{0pt}
    \setlength{\parskip}{6pt plus 2pt minus 1pt}}
}{% if KOMA class
  \KOMAoptions{parskip=half}}
\makeatother
\usepackage{xcolor}
\usepackage{longtable,booktabs,array}
\usepackage{calc} % for calculating minipage widths
\usepackage{etoolbox}
\makeatletter
\patchcmd\longtable{\par}{\if@noskipsec\mbox{}\fi\par}{}{}
\makeatother
\IfFileExists{footnotehyper.sty}{\usepackage{footnotehyper}}{\usepackage{footnote}}
\makesavenoteenv{longtable}
\usepackage{graphicx}
\makeatletter
\def\maxwidth{\ifdim\Gin@nat@width>\linewidth\linewidth\else\Gin@nat@width\fi}
\def\maxheight{\ifdim\Gin@nat@height>\textheight\textheight\else\Gin@nat@height\fi}
\makeatother
\setkeys{Gin}{width=\maxwidth,height=\maxheight,keepaspectratio}
\makeatletter
\def\fps@figure{htbp}
\makeatother
\providecommand{\tightlist}{%
  \setlength{\itemsep}{0pt}\setlength{\parskip}{0pt}}
\newlength{\cslhangindent}
\newlength{\csllabelwidth}
\newlength{\cslentryspacingunit} % times entry-spacing
\newenvironment{CSLReferences}[2] % #1 hanging-ident, #2 entry spacing
 {% don't indent paragraphs
  \setlength{\parindent}{0pt}
  \ifodd #1
  \let\oldpar\par
  \def\par{\hangindent=\cslhangindent\oldpar}
  \fi
  \setlength{\parskip}{#2\cslentryspacingunit}
 }%
 {}
\usepackage{calc}

\ifLuaTeX
\usepackage[bidi=basic]{babel}
\else
\usepackage[bidi=default]{babel}
\fi
\babelprovide[main,import]{american}
\def\languageshorthands#1{}
\ifLuaTeX
  \usepackage{selnolig}  % disable illegal ligatures
\fi
\IfFileExists{bookmark.sty}{\usepackage{bookmark}}{\usepackage{hyperref}}
\IfFileExists{xurl.sty}{\usepackage{xurl}}{} % add URL line breaks if available
\hypersetup{
  pdftitle={HoverFast: an accurate, high-throughput, clinically
deployable nuclear segmentation tool for brightfield digital pathology
images},
  pdfauthor={Petros Liakopoulos, Julien Massonnet, Jonatan Bonjour,
Medya Tekes Mizrakli, Simon Graham, Michel A. Cuendet, Amanda H. Seipel,
Olivier Michielin, Doron Merkler, Andrew Janowczyk},
  pdflang={en-US},
  colorlinks=true,
  linkcolor={Maroon},
  filecolor={Maroon},
  citecolor={Blue},
  urlcolor={Blue},
  pdfcreator={LaTeX via pandoc}}

\title{HoverFast: an accurate, high-throughput, clinically deployable
nuclear segmentation tool for brightfield digital pathology images}

\usepackage[affil-it]{authblk}
\usepackage{orcidlink}
\author[1%
  *%
  \ensuremath\mathparagraph]{Petros Liakopoulos%
    \,\orcidlink{0009-0005-2015-6795}\,%
    }
\author[2%
  *%
  ]{Julien Massonnet%
    \,\orcidlink{0009-0004-9515-6100}\,%
    }
\author[1%
  ]{Jonatan Bonjour%
    \,\orcidlink{0009-0006-8165-6897}\,%
    }
\author[3%
  ]{Medya Tekes Mizrakli%
    }
\author[4%
  ]{Simon Graham%
    }
\author[1%
  ]{Michel A. Cuendet%
    }
\author[2%
  ]{Amanda H. Seipel%
    }
\author[1%
  ]{Olivier Michielin%
    }
\author[2%
  ]{Doron Merkler%
    }
\author[1,2,5%
  ]{Andrew Janowczyk%
    }

\affil[1]{Service of Precision Oncology, Department of Oncology,
University of Geneva and Geneva University Hospitals, Geneva,
Switzerland}
\affil[2]{Division of Clinical Pathology, , Departments of Pathology and
Immunology \& Diagnostics, University of Geneva and Geneva University
Hospitals, Geneva, Switzerland}
\affil[3]{Section of Communication Systems, School of Computer and
Communication Sciences, École Polytechnique Fédérale de Lausanne,
Lausanne, Switzerland.}
\affil[4]{Histofy Ltd, Birmingham, United Kingdom}
\affil[5]{Emory University, Atlanta, GA, USA.}
\affil[$\mathparagraph$]{Corresponding author}
\affil[*]{These authors contributed equally.}
\begin{document}
\maketitle

\hypertarget{summary}{%
\section{Summary}\label{summary}}

In computational digital pathology, accurate nuclear segmentation of
Hematoxylin and Eosin (H\&E) stained whole slide images (WSIs) is a
critical step for many analyses and tissue characterizations. One
popular deep learning-based nuclear segmentation approach, HoverNet
(\protect\hyperlink{ref-graham2019hover}{Graham et al., 2019}), offers
remarkably accurate results but lacks the high-throughput performance
needed for clinical deployment in resource-constrained settings. Our
approach, HoverFast, aims to provide fast and accurate nuclear
segmentation in H\&E images using knowledge distillation from HoverNet.
By redesigning the tool with software engineering best practices,
HoverFast introduces advanced parallel processing capabilities,
efficient data handling, and optimized postprocessing. These
improvements facilitate scalable high-throughput performance, making
HoverFast more suitable for real-time analysis and application in
resource-limited environments. Using a consumer grade Nvidia A5000 GPU,
HoverFast showed a 21x speed improvement as compared to HoverNet;
reducing mean analysis time for 40x WSIs from \textasciitilde2 hours to
6 minutes while retaining a concordant mean Dice score of 0.91 against
the original HoverNet output. Peak memory usage was also reduced 71\%
from 44.4GB, to 12.8GB, without requiring SSD-based caching. To ease
adoption in research and clinical contexts, HoverFast aligns with
best-practices in terms of (a) installation, and (b) containerization,
while (c) providing outputs compatible with existing popular open-source
image viewing tools such as QuPath
(\protect\hyperlink{ref-bankhead2017qupath}{Bankhead et al., 2017}).
HoverFast has been made open-source and is available at
\url{andrewjanowczyk.com/open-source-tools/hoverfast}.

\hypertarget{statement-of-need}{%
\section{Statement of need}\label{statement-of-need}}

The increasing popularity of digitized pathology images in both research
and clinical practice has spurred the widespread adoption of deep
learning (DL) approaches for automating various tasks, with nuclear
segmentation standing out as a crucial step in many analyses. This
segmentation process involves delineating the contours of cell nuclei
within a 2D whole slide image (WSI). Nuclei, rather than complete cells,
are targeted due to strong contrast afforded by routinely employed
hematoxylin staining. Hematoxylin's selective affinity for nucleic acids
results in the distinct visualization of nuclei in purple, facilitating
their clear identification amidst less prominently stained cytoplasm and
other cellular constituents. Given the small size of nuclei, their
segmentation typically takes place at 40x magnification
(\textasciitilde0.25 microns per pixel (mpp)); the highest magnification
supported by most current digital slide scanners. Working at this scale
can be time-intensive for algorithms, especially on consumer grade GPUs,
as WSIs are especially large, reaching up to 120,000x120,000 pixels.
While several existing tools like StarDist
(\protect\hyperlink{ref-schmidt2018}{Schmidt et al., 2018})
(\protect\hyperlink{ref-weigert2020}{Weigert et al., 2020}) and CellPose
(\protect\hyperlink{ref-stringer2021cellpose}{Stringer et al., 2021})
have been developed to tackle the challenge of nuclear segmentation,
HoverNet(\protect\hyperlink{ref-graham2019hover}{Graham et al., 2019})
has emerged as one of the leading solutions in terms of segmentation
accuracy, particularly for its application to H\&E-stained tissue.

Despite its accurate results, HoverNet remains resource-intensive and
time-consuming due to its high model parameter count and lengthy
post-processing steps. HoverNet additionally requires significant SSD
storage for caching during runtime, often reaching over 120GB per WSI.
These properties make it challenging to deploy in more resource limited
settings such as consumer grade workstations or in clinical environments
requiring high-throughput processing. Therefore, there is an emerging
need for a fast, accurate, and computationally efficient tool that can
make large-scale nuclear segmentation more accessible for both research
and clinical applications.

Motivated by the need for accurate yet efficient nuclear segmentation,
we introduce HoverFast. This tool replicates the output of the
established HoverNet model while achieving superior computational
efficiency. HoverFast achieves this through knowledge distillation, a
technique where a smaller ``student'' model (HoverFast) learns to
capture the knowledge from a larger ``teacher'' model (HoverNet). The
goal is to enable the student model to achieve comparable performance to
that of the teacher model, while requiring significantly less
computational resources for inference
(\protect\hyperlink{ref-hinton2015distilling}{Hinton et al., 2015}),
(\protect\hyperlink{ref-hu2022teacher}{Hu et al., 2022}).

To facilitate the knowledge distillation process, HoverFast presents a
training pipeline, using HoverNet output as ground truth (see
\textbf{Figure 1}), that enables the resulting model to have 30 times
fewer parameters. As implemented, HoverFast provides:

\begin{itemize}
\tightlist
\item
  A containerized docker script to generate HoverNet ground truth on
  user-provided data
\item
  A training pipeline for a custom HoverFast model
\item
  Alternatively, a pre-trained cross-organ model for inference
\item
  An inference pipeline for tiles and WSIs with tissue masks to
  delineate area of computation
\item
  Compressed JSON output file directly compatible with QuPath
\item
  A speedup of 21x over HoverNet, on consumer-grade compute
  infrastructure
\end{itemize}

\begin{figure}
\centering
\includegraphics[width=\textwidth,height=2.91667in]{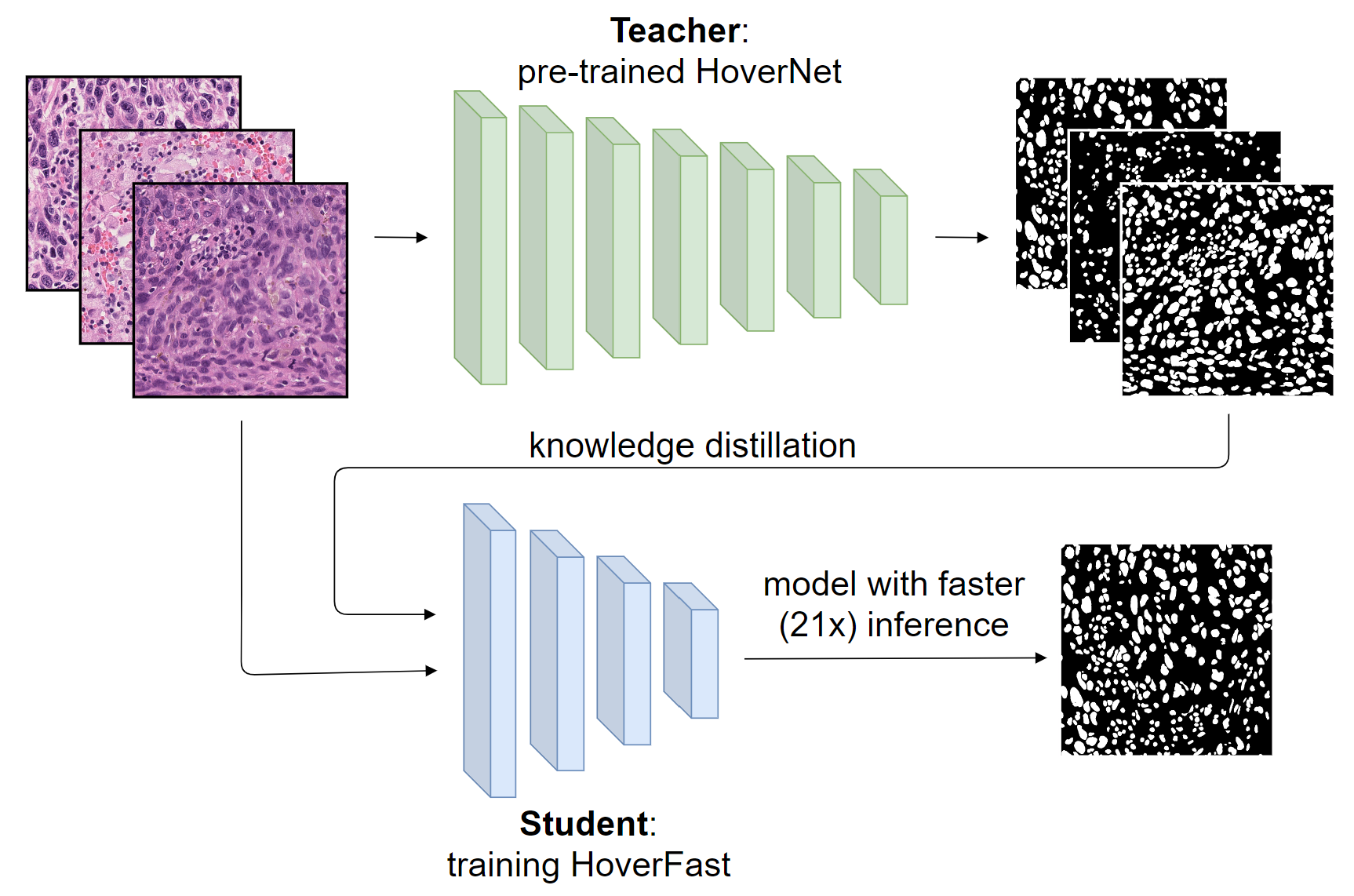}
\caption{Overview of training pipeline for HoverFast. H\&E tiles are
segmented using HoverNet. The nuclei masks, as well as the original H\&E
tiles, are passed to the smaller HoverFast architecture for training.
The resulting HoverFast model provides a highly optimized inference and
post-processing framework that can then be used for nuclear segmentation
on WSIs with a 21x speed improvement over HoverNet.\label{fig:Figure 1}}
\end{figure}

\hypertarget{implementation}{%
\section{Implementation}\label{implementation}}

\hypertarget{inference}{%
\subsection{Inference}\label{inference}}

HoverFast has a command-line interface (CLI) written in Python 3.11 and
utilizes the PyTorch framework
(\protect\hyperlink{ref-paszke2017automatic}{Paszke et al., 2017}). We
replicated the structure of the HoverNet model as described by Graham et
al (\protect\hyperlink{ref-graham2019hover}{Graham et al., 2019})
without the nuclear classification branch. For the backbone, we used a
modified 940k parameter Multi-scale UNet
(\protect\hyperlink{ref-su2021msu}{Su et al., 2021}) in place of
HoverNet's 33.6 million parameter ResNet50, yielding a reduction in
model parameter count by a factor of 30 (see \textbf{Appendix 1}).

HoverFast's post-processing pipeline was heavily optimized using
scikit-learn's (\protect\hyperlink{ref-scikit-learn}{Pedregosa et al.,
2011}) regionprops and watershed functions to effectively identify and
split merged cells. To improve throughput after batch model inference,
regions are processed in parallel using a ``multi-worker, single
writer'' approach. This involves each worker independently (a)
post-processing its assigned region, and then (b) generating nuclei
polygon coordinates using OpenCV
(\protect\hyperlink{ref-opencv_library}{Bradski, 2000}), before (c)
sending to the single writing process for saving as a QuPath
(\protect\hyperlink{ref-bankhead2017qupath}{Bankhead et al., 2017})
compatible gzip-compressed JSON file. A Docker and Singularity container
of HoverFast are provided.

\hypertarget{training}{%
\subsection{Training}\label{training}}

To help users train their own models, we provide a Docker container with
HoverNet installed, and a script that (a) accepts a directory of WSIs
(or tiles), (b) randomly extracts a user-specified number of tiles, (c)
employs HoverNet on these tiles to generate labeled masks of nuclei, and
finally (d) saves the original images and associated masks into two
PyTables files, one for training and one for validation. HoverFast can
then accept these PyTables files as arguments in its training script to
yield a use-case specific model. We employ knowledge distillation during
HoverFast training using the same loss function as HoverNet. The teacher
model, HoverNet, guides the student model, HoverFast, by providing the
ground truth binary mask as target. Additionally, HoverFast learns to
reproduce the horizontal and vertical distance maps allowing HoverFast
to inherit HoverNet's post-processing abilities for separating touching
nuclei.

\hypertarget{experiments}{%
\section{Experiments}\label{experiments}}

\hypertarget{experiment-1-comparison-of-hovernet-and-hoverfast-on-a-cross-organ-dataset}{%
\subsection{Experiment 1: Comparison of HoverNet and HoverFast on a
cross organ
dataset}\label{experiment-1-comparison-of-hovernet-and-hoverfast-on-a-cross-organ-dataset}}

Employing n=97 WSIs of diverse tissue types, from The Cancer Genome
Atlas (TCGA) (\protect\hyperlink{ref-weinstein2013cancer}{Weinstein et
al., 2013}), 15 randomly selected tiles of 1,024x1,024 pixels from each
WSI were extracted. A HoverFast model was then trained as described in
\protect\hyperlink{training}{Training}, for 100 epochs with a batch size
of 16. For validation, 74 tiles of 1,024x1,024 pixels from 14 slides of
diverse tissue types were used. Inference using both HoverNet and
HoverFast was performed on the validation tiles, and the binary masks of
predicted cell nuclei were overlapped to obtain a Dice score between the
two tools. The resulting Dice score of 0.91 appears concordant with the
qualitative results (see \textbf{Figure 2}); these show very similar
segmentation results, with HoverFast able to segment slightly more faint
nuclei than HoverNet.

\begin{figure}
\centering
\includegraphics[width=\textwidth,height=4.16667in]{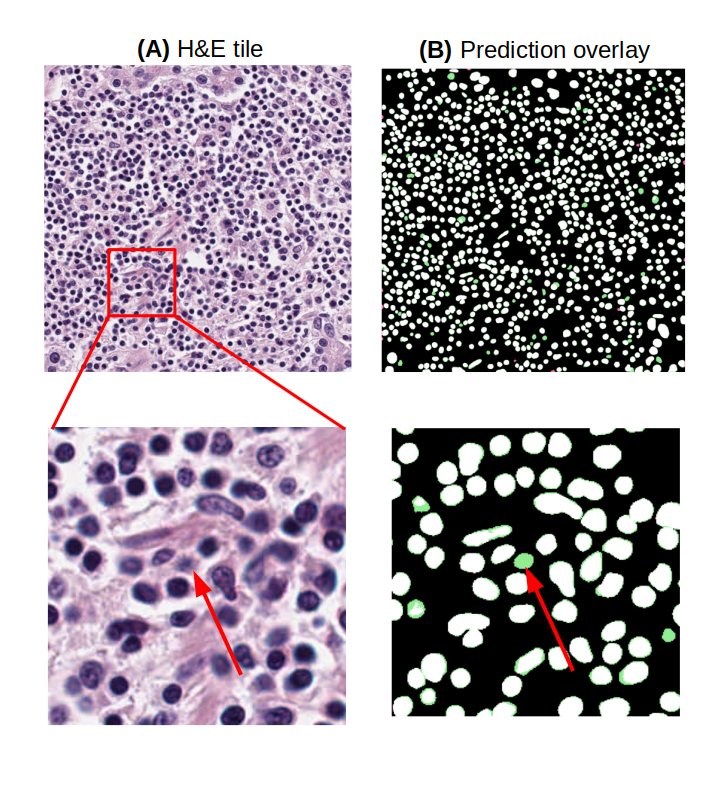}
\caption{(A) 1,024x1,024 tile of H\&E tissue. (B) Overlay of binary
masks. Pixels in white are predicted by both models, with those in
green/pink predicted only by HoverFast/HoverNet, respectively. For this
patch, the Dice score is 0.91. Here we can generally see that while
HoverFast detects more nuclei, those predictions remain reasonable and
clearly visible on H\&E, suggesting a potentially higher quality
result.\label{fig:Figure 2}}
\end{figure}

\hypertarget{experiment-2-comparison-of-hovernet-cross-tissue-hoverfast-and-site-specific-hoverfast}{%
\subsection{Experiment 2: Comparison of HoverNet, cross-tissue
HoverFast, and site-specific
HoverFast}\label{experiment-2-comparison-of-hovernet-cross-tissue-hoverfast-and-site-specific-hoverfast}}

From n=54 melanoma samples, (a) for training: 20 1,024x1,024 tiles were
randomly selected per slide from within available tumor masks, and (b)
for validation: 50 tiles of 1,024x1,024 from 8 slides were selected. For
evaluation, 3 models were compared: (i) HoverNet, as a baseline, (ii)
the HoverFast model trained in
\protect\hyperlink{experiment-1-comparison-of-hovernet-and-hoverfast-on-a-cross-organ-dataset}{Experiment
1}, and (iii) a melanoma specific model trained following the procedure
in \protect\hyperlink{training}{Training}. The HoverFast models had Dice
scores of 0.88 and 0.91 respectively against HoverNet, with qualitative
results indicating a high degree of similarity. There were slight
changes on nuclei edges and faint nuclei (see \textbf{Figure 3}), with a
systematic superiority for the tissue-specific output. Taken together,
the increased accuracy in the melanoma specific model demonstrates that
investing in training a dataset-specific model appears to provide added
value.

\begin{figure}
\centering
\includegraphics{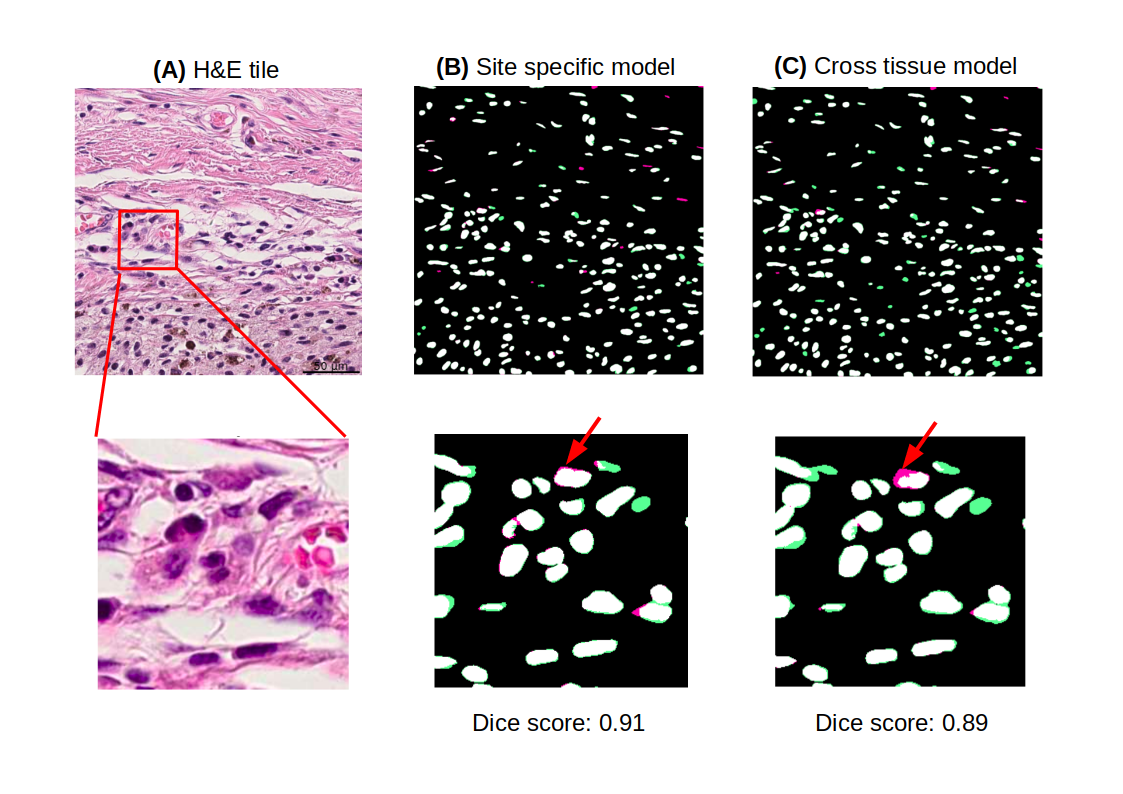}
\caption{(A) 1,024x1,024 tile of H\&E tissue (top) with a higher
magnification region of interest (bottom). (B) Overlay of binary masks
for the single tissue model. Pixels in white are predicted by both
models, with those in green/pink predicted only by HoverFast/HoverNet,
respectively. (C) Overlay of binary mask for the cross-tissue model. The
tissue specific model shows a closer resemblance to HoverNet in terms of
cell outlines.\label{fig:Figure 3}}
\end{figure}

\hypertarget{experiment-3-benchmarks-comparing-processing-time-and-memory-footprint}{%
\subsection{Experiment 3: Benchmarks comparing processing time and
memory
footprint}\label{experiment-3-benchmarks-comparing-processing-time-and-memory-footprint}}

To compare computational speed, n=4 slides from TCGA with corresponding
tissue masks generated with HistoQC
(\protect\hyperlink{ref-janowczyk2019histoqc}{Janowczyk et al., 2019})
were analyzed on a machine with a 16 core Intel(R) Core(TM) i9-12900K
CPU, a Nvidia A5000 GPU with 24GB of VRAM, and 128Gb of DDR5 RAM. For
both HoverNet and HoverFast, the GPU batch size was set to maximize GPU
memory usage. For HoverNet, a batch size of 90 was used, with 20 CPU
threads for pre- and post-processing. Similarly, for HoverFast, a batch
size of 11 and 20 CPU threads were used. A mean speed improvement of
20.8x times (see \textbf{Table 1}) was demonstrated. The maximum RAM
consumption was reduced by 71\% with 44.4 GB for HoverNet versus 12.8 GB
for HoverFast. Additionally, HoverNet required a peak of 118 GB of SSD
space for its cache during run-time, while HoverFast did not appear to
require any.

\begin{longtable}[]{@{}llll@{}}
\caption{Detailed table of computation time per slide for each tool with
associated speedup {}}\tabularnewline
\toprule\noalign{}
\emph{Slide ID} & \emph{HoverNet} & \emph{HoverFast} & \emph{Speedup} \\
\midrule\noalign{}
\endfirsthead
\toprule\noalign{}
\emph{Slide ID} & \emph{HoverNet} & \emph{HoverFast} & \emph{Speedup} \\
\midrule\noalign{}
\endhead
\bottomrule\noalign{}
\endlastfoot
Slide 1 & 58mins 5s & 3mins 1s & 19.2x \\
Slide 2 & 1hr 11mins 38s & 3mins 33s & 20.2x \\
Slide 3 & 2hrs 55mins 24s & 8mins 4s & 21.7x \\
Slide 4 & 3hrs 4mins 25s & 8mins 50s & 20.9x \\
\textbf{Total} & \textbf{8hrs 9mins 32s} & \textbf{23mins 28s} &
\textbf{20.8x} \\
\end{longtable}

\hypertarget{discussion-and-conclusions}{%
\section{Discussion and Conclusions:}\label{discussion-and-conclusions}}

HoverFast represents a practical solution to the challenge of nuclear
segmentation in WSIs, emphasizing speed, resource efficiency and local
trainability. It distinguishes itself by providing a significant speedup
in processing time with a 21x improvement over HoverNet on consumer
grade hardware in addition to a more than 3x reduction in RAM footprint
while also eliminating hard-drive based caching. This efficiency is
crucial for users with limited resources, enabling faster analysis while
retaining segmentation results highly comparable to those of HoverNet.
While a pre-trained cross-tissue model is provided with the software, if
higher accuracy and greater similarity to HoverNet is required, a cohort
specific model should be trained. Additionally, although HoverFast does
have a built-in feature for tissue detection, we highly recommend the
use of quality control tools, such as HistoQC to obtain more robust
tissue masks, thus avoiding computation on artefactual regions and
further reducing computation time. HoverFast is easy to install and
provides simple drag and drop output compatibility with QuPath. It is
publicly available for use and modification at \url{andrewjanowczyk.com/open-source-tools/hoverfast}.

\hypertarget{appendix-1}{%
\subsection{Appendix 1}\label{appendix-1}}

\includegraphics[width=\textwidth,height=5.55556in]{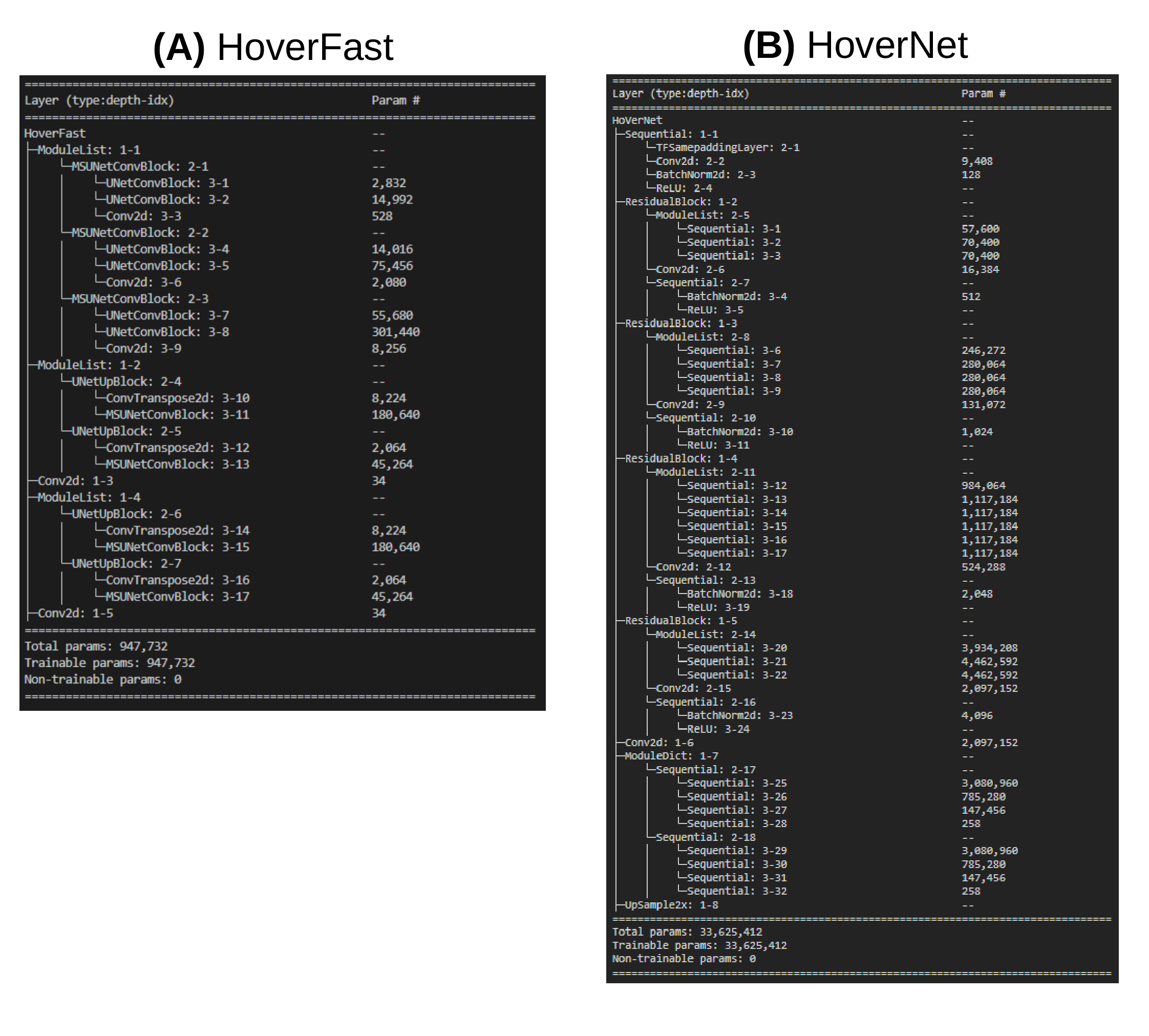}

The appendix presents a comparison between HoverFast and HoverNet
architectures. This information was produced using the Python Torchinfo
Summary package. The first column outlines the model architecture, while
the second delves into the number of parameters for each layer. It is
noteworthy that HoverFast substantially smaller number of parameters
(roughly 30 times fewer than HoverNet) which, along with the optimized
post-processing and file handling, translates to lower memory footprint
and faster processing time. This enables HoverFast to handle larger
batches of data and allowing parallel post-processing computation,
ultimately leading to a well-suited tool for resource limited
environments.

\hypertarget{references}{%
\section*{References}\label{references}}
\addcontentsline{toc}{section}{References}

\hypertarget{refs}{}
\begin{CSLReferences}{1}{0}
\leavevmode\vadjust pre{\hypertarget{ref-bankhead2017qupath}{}}%
Bankhead, P., Loughrey, M. B., Fernández, J. A., Dombrowski, Y., McArt,
D. G., Dunne, P. D., McQuaid, S., Gray, R. T., Murray, L. J., Coleman,
H. G., \& others. (2017). QuPath: Open source software for digital
pathology image analysis. \emph{Scientific Reports}, \emph{7}(1), 1--7.

\leavevmode\vadjust pre{\hypertarget{ref-opencv_library}{}}%
Bradski, G. (2000). {The OpenCV Library}. \emph{Dr. Dobb's Journal of
Software Tools}.

\leavevmode\vadjust pre{\hypertarget{ref-graham2019hover}{}}%
Graham, S., Vu, Q. D., Raza, S. E. A., Azam, A., Tsang, Y. W., Kwak, J.
T., \& Rajpoot, N. (2019). Hover-net: Simultaneous segmentation and
classification of nuclei in multi-tissue histology images. \emph{Medical
Image Analysis}, \emph{58}, 101563.

\leavevmode\vadjust pre{\hypertarget{ref-hinton2015distilling}{}}%
Hinton, G., Vinyals, O., \& Dean, J. (2015). Distilling the knowledge in
a neural network. \emph{arXiv Preprint arXiv:1503.02531}.

\leavevmode\vadjust pre{\hypertarget{ref-hu2022teacher}{}}%
Hu, C., Li, X., Liu, D., Chen, X., Wang, J., \& Liu, X. (2022).
Teacher-student architecture for knowledge learning: A survey.
\emph{arXiv Preprint arXiv:2210.17332}.

\leavevmode\vadjust pre{\hypertarget{ref-janowczyk2019histoqc}{}}%
Janowczyk, A., Zuo, R., Gilmore, H., Feldman, M., \& Madabhushi, A.
(2019). HistoQC: An open-source quality control tool for digital
pathology slides. \emph{JCO Clinical Cancer Informatics}, \emph{3},
1--7.

\leavevmode\vadjust pre{\hypertarget{ref-paszke2017automatic}{}}%
Paszke, A., Gross, S., Chintala, S., Chanan, G., Yang, E., DeVito, Z.,
Lin, Z., Desmaison, A., Antiga, L., \& Lerer, A. (2017). \emph{Automatic
differentiation in PyTorch}.

\leavevmode\vadjust pre{\hypertarget{ref-scikit-learn}{}}%
Pedregosa, F., Varoquaux, G., Gramfort, A., Michel, V., Thirion, B.,
Grisel, O., Blondel, M., Prettenhofer, P., Weiss, R., Dubourg, V.,
Vanderplas, J., Passos, A., Cournapeau, D., Brucher, M., Perrot, M., \&
Duchesnay, E. (2011). Scikit-learn: Machine learning in python.
\emph{Journal of Machine Learning Research}, \emph{12}, 2825--2830.

\leavevmode\vadjust pre{\hypertarget{ref-schmidt2018}{}}%
Schmidt, U., Weigert, M., Broaddus, C., \& Myers, G. (2018). Cell
detection with star-convex polygons. \emph{Medical Image Computing and
Computer Assisted Intervention - {MICCAI} 2018 - 21st International
Conference, Granada, Spain, September 16-20, 2018, Proceedings, Part
{II}}, 265--273. \url{https://doi.org/10.1007/978-3-030-00934-2_30}

\leavevmode\vadjust pre{\hypertarget{ref-stringer2021cellpose}{}}%
Stringer, C., Wang, T., Michaelos, M., \& Pachitariu, M. (2021).
Cellpose: A generalist algorithm for cellular segmentation. \emph{Nature
Methods}, \emph{18}(1), 100--106.

\leavevmode\vadjust pre{\hypertarget{ref-su2021msu}{}}%
Su, R., Zhang, D., Liu, J., \& Cheng, C. (2021). MSU-net: Multi-scale
u-net for 2D medical image segmentation. \emph{Frontiers in Genetics},
\emph{12}, 639930.

\leavevmode\vadjust pre{\hypertarget{ref-weigert2020}{}}%
Weigert, M., Schmidt, U., Haase, R., Sugawara, K., \& Myers, G. (2020).
Star-convex polyhedra for 3D object detection and segmentation in
microscopy. \emph{The IEEE Winter Conference on Applications of Computer
Vision (WACV)}. \url{https://doi.org/10.1109/WACV45572.2020.9093435}

\leavevmode\vadjust pre{\hypertarget{ref-weinstein2013cancer}{}}%
Weinstein, J. N., Collisson, E. A., Mills, G. B., Shaw, K. R.,
Ozenberger, B. A., Ellrott, K., Shmulevich, I., Sander, C., \& Stuart,
J. M. (2013). The cancer genome atlas pan-cancer analysis project.
\emph{Nature Genetics}, \emph{45}(10), 1113--1120.

\end{CSLReferences}

\end{document}